\documentclass{article}
\makeatletter
\@addtoreset{equation}{section}

\makeatother

\begin{document}

\begin{flushright}
Oct 2002

KEK-TH-849
\end{flushright}

\begin{center}

\vspace{5cm}

{\Large Closed String Tachyons and RG flows}

\vspace{2cm}

Takao Suyama \footnote{e-mail address : tsuyama@post.kek.jp}

\vspace{1cm}

{\it Theory Group, KEK}

{\it Tsukuba, Ibaraki 305-0801, Japan}

\vspace{4cm}

{\bf Abstract} 

\end{center}

We consider bulk tachyon condensations in a non-linear sigma model whose low energy effective theory contains 
a nontrivial scalar potential. 
We argue that one would typically encounter a strong coupling background along a RG flow corresponding to 
a bulk tachyon condensation, beyond which the RG analysis would not be reliable. 
In a range of the flow in which the string coupling constant is small, we can show that the tachyon condensation 
actually decreases the central charge of the sigma model. 

\newpage

\section{Introduction}

\vspace{5mm}

Tachyon condensation is a good arena to consider the off-shell dynamics of string theory. 
The string perturbation theory can provide us with only on-shell quantities, and thus one has to employ other 
ingredients to investigate that phenomenon. 
The use of string field theory is the most straightforward way to investigate off-shell properties of 
string theory, 
and in fact it has appeared to be a very powerful 
tool for studying open string tachyon condensations \cite{SFT}\cite{SFT2}. 
In addition, there is another framework which is useful in some cases. 
This is the renormalization group analysis of worldsheet theories \cite{RG}. 
According to this, one can perform an explicit calculation to obtain the exact value of a physical quantity, 
the tension of a D-brane \cite{BSFT}\cite{BSFT2}. 

Intuitively, what one expects to occur in tachyon condensations would be decays of unstable {\it on-shell} 
backgrounds (or states) into other stable or less unstable {\it on-shell} backgrounds (or states). 
In the case of open string tachyon condensations, tachyons appear on unstable D-brane systems, and they decay 
into the closed string vacuum or lower-dimensional D-brane systems. 
In the localized closed string tachyon condensations, it has been discussed that the Melvin background or twisted 
circle geometry would decay into the flat spacetime \cite{Melvin}\cite{Melvin2}, 
and non-supersymmetric orbifolds would also decay into the 
flat spacetime \cite{APS}\cite{Vafa}\cite{HKMM}.  
There are also some conjectures on various non-supersymmetric string theories, which are based on relations to 
non-supersymmetric backgrounds of superstring theories \cite{TypeII}\cite{hetero}. 

However, such an intuition seems to conflict with the results of the RG analysis in the condensation of tachyons 
which can propagate in the whole target spacetime. 
Every on-shell backgrounds provide the corresponding worldsheet theories whose central charges are 
equal to a specific  
value; 26 for bosonic strings and 15 for superstrings. 
On the other hand, bulk tachyon condensations are considered to decrease the central charge of the corresponding 
worldsheet theory. 
Therefore, bulk tachyon condensations which connect two on-shell backgrounds seem to contradict the c-theorem 
\cite{c-th}. 

In this paper, we try to give some comments on the bulk tachyon condensations, and argue that the physical 
intuition discussed above does not conflict with c-theorem. 
The point is that the worldsheet analysis is the tree-level approximation of string theory, which is 
valid when the string coupling is small everywhere in the target spacetime. 
We will discuss that one would typically encounter a strong coupling background along the RG flow. 
In addition, we will show that bulk tachyon condensations actually decrease the central charge in a range of the 
RG flow in which the string coupling is a small constant. 

This paper is organized as follows. 
In section \ref{review}, we review the worldsheet approach to tachyon condensations. 
The low energy effective theories of worldsheet theories and their classical solutions are examined in section 
\ref{LEET}. 
The central charge is discussed in relation to the c-theorem in section \ref{c}. 
Section \ref{discussion} is devoted to discussions.

\vspace{1cm}

\section{Renormalization group analysis}   \label{review}

\vspace{5mm}

The analysis of tachyon condensations would require off-shell information of string theory. 
For example, one usually considers a zero-momentum tachyon, which is certainly not an on-shell state, to obtain 
a potential for the tachyon. 
Therefore, the most efficient ingredient for such analyses would be string field theory 
\cite{SFT}\cite{SFT2}. 
However, there is an equally useful tool for the problem. 
It is the worldsheet theory combined with the renormalization group analysis \cite{RG}. 

In worldsheet theories, a condensation of a state is described by the addition of the corresponding vertex operator 
to the worldsheet action. 
In the case of open string states, the vertex operator is inserted at the boundary of the worldsheet. 
When the state which condenses is a zero-momentum tachyon, the term added breaks the conformal invariance, and 
the term induces a RG flow since it is a relevant deformation of the initial CFT. 
A prescription to read off the off-shell dynamics of string theory corresponding to a tachyon condensation is to 
relate the RG flow to the physical process \cite{RG}. 
That is, if there is a RG flow induced by a relevant deformation which connects two CFT's, then it is interpreted 
that the UV CFT decays into the IR CFT via a tachyon condensation. 
Here UV(IR) CFT means the UV(IR) limit of the RG flow. 

This prescription is successfully applied to the case of open string tachyons \cite{RG}\cite{BSFT}\cite{BSFT2}. 
In this case, open string tachyon condensations are described by deformations of boundary conditions of 
worldsheet fields, while the bulk of the worldsheet does not affected. 
Thus the analysis of such deformations would be easier than that of deformations of the bulk CFT. 
Moreover, there is a simple deformation by which one can explicitly show the RG flow and its IR limit, for 
example, 
\begin{equation}
S = \frac1{4\pi\alpha'}\int_{D_2} d^2\sigma\ \partial_\alpha X\partial_\alpha X
   +\int_{\partial D_2}d\theta (a+uX^2), 
\end{equation}
for bosonic string theory \cite{X^2}. 
The result of this deformation is just a change of the boundary condition of $X$ from Neumann to 
Dirichlet. 
In the spacetime point of view, this indicates a decay of a D-brane to a lower dimensional D-brane. 
This phenomenon can also be described by the effective field theory on the D-brane. 
The exact form of the effective action up to two derivatives can be obtained from the partition function of the 
above worldsheet theory \cite{BSFT}\cite{BSFT2}. 

The idea of applications of the renormalization group seems to be useful even in a class of closed string tachyon 
condensations. 
The presence of tachyons implies that there is no preserved spacetime supersymmetry, and thus it is difficult to 
control quantum corrections. 
However, even in such cases, it is possible to keep the worldsheet supersymmetry, and it enables one to control 
quantum corrections of the worldsheet theory if the number of supercharges is large enough. 
The research in this direction has been performed, and claimed that, for example, various non-compact orbifolds 
would decay into the flat spacetime or supersymmetric orbifolds \cite{Vafa}\cite{HKMM}. 
At first sight, those claims seem to contradict the well-known c-theorem. 
There are arguments which point out possibilities for the c-theorem not to be true \cite{APS}\cite{HKMM}. 

\vspace{2mm}

What the RG analysis does seems to be similar to a treatment of off-shell backgrounds in ordinary field theories. 
Suppose that one expands a theory around an off-shell background. 
Then the tadpole is produced. 
In such a case, one can achieve the theory expanded around the correct background or vacuum by including the 
effects of the tadpoles, i.e. summing up the tadpoles. 
Then the summation shifts the background to the other background in which the tadpole vanishes. 
On the other hand, the RG analysis of the tachyon insertion studies, roughly speaking, what happens when the 
effect of the insertion becomes important. 
Therefore, the relation between the RG flows and tachyon condensations is intuitively very natural. 

One difference is that, in field theories, one has to include loop corrections to achieve the correct vacuum, 
while the worldsheet theory is the classical theory of strings and no loop corrections are included. 
Of course, the worldsheet theory on the sphere is an approximation of string theory, which is valid when the 
string coupling constant is small. 
Thus, under that condition, the RG analysis and the field theory analysis seem to be equivalent to each other. 

\vspace{2mm}

One would not be able to evade the c-theorem in the case of closed string tachyons propagating in the bulk 
spacetime. 
A naive expectation for the fate of such theories is that the endpoints of the condensations would be 
non-critical string theories, since the central charge must decrease along the RG flows. 
In the following sections, we will consider such situations in detail, 
focusing on the relation between worldsheet theories 
and their spacetime effective theories.

\vspace{1cm}

\section{Low energy effective theories}    \label{LEET}

\vspace{5mm}

Consider a simple scalar field theory
\begin{equation}
S = \int d^4x\ \left[ -\frac12\partial_\mu\phi\partial^\mu\phi-\frac12(\phi^2-\phi_0^2)^2 \right].
\end{equation}
The scalar $\phi$ is tachyonic with $m_\phi^2=-2\phi_0^2$, and thus the vacuum $\phi=0$ is unstable. 
The stable vacua are $\phi=\pm\phi_0$. 
The fluctuation $\varphi$ around the stable vacua is massive with $m_\varphi^2=4\phi_0^2$. 
Note that the tachyon $\phi$ can propagate in the whole bulk spacetime, but there is no difficulty in considering 
its condensation which shifts the vev of $\phi$ to the correct values. 
The situation would not change drastically when the above theory is coupled to gravity. 
The notion of stability may, however, be modified 
due to effects of curved backgrounds, like in the case of AdS space 
\cite{AdS}. 
Nevertheless, it would be natural to expect that, in theories with nontrivial scalar potentials, the vevs of 
scalars corresponding to a potential maximum or a saddle are shifted to the vevs corresponding to a potential 
minimum, via a condensation of the scalars, and the vacuum is stabilized. 

In the point of view of the worldsheet RG, however, 
it seems difficult to understand the above-mentioned situations. 
Suppose that a low energy effective theory of a sigma model has a nontrivial potential. 
When the theory is expanded around a potential maximum or a saddle, the theory contains in its spectrum a bulk 
tachyon or a massless particle whose vev destabilizes the vacuum. 
One would expect that turning on a vev of the tachyon (the massless particle) corresponds in the worldsheet 
theory to a relevant (marginally relevant) deformation. 
According to the c-theorem, the central charge of the worldsheet theory decreases as the vev goes down the 
potential hill. 
Therefore, the worldsheet theory corresponding to a potential minimum in the vicinity of the initial extremum 
should have a central charge which is 
smaller than that of the initial worldsheet theory. 
However, if the above two extrema correspond to classical solutions of the low energy effective theory, 
both central charges must be 26 for bosonic string theory and 15 for superstring theories. 
Thus bulk tachyon condensations which connect two classical solutions of the effective theory seem to contradict 
the c-theorem, even if the solutions are connected by shifting the scalar vevs continuously. 

\vspace{2mm}

This puzzling situation can be understood by examining the validity of the 
use of the worldsheet RG for the problems. 
The important point is that the worldsheet theory is the tree-level approximation of string theory, 
and it is useful only when 
the string coupling is small everywhere in the target spacetime. 
If a worldsheet theory corresponds to a strong coupling spacetime background, the tree-level analysis would get 
large quantum corrections. 

\vspace{3mm}

For definiteness, let us consider classical solutions of the action 
\begin{eqnarray}
S &=& \frac1{2\kappa^2}\int d^Dx\sqrt{-g}e^{-2\Phi}\left[ R+4\partial_\mu\Phi\partial^\mu\Phi
     -\frac1{12}H_{\mu\nu\rho}H^{\mu\nu\rho}\right. \nonumber \\
  & & \hspace*{2cm}\left.\frac{ }{ }
     -f_{ab}(\phi)F^a_{\mu\nu}F^{b\mu\nu}-g_{IJ}(\phi)D_\mu\phi^I D^\mu\phi^J-V(\phi)\right]. 
            \label{gauged}
\end{eqnarray}
The action of this form can be obtained, for example, by compactifying the heterotic strings on a torus with 
flux \cite{SUGRA}. 
The scalars $\phi^I$ may couple to the gauge fields $A_\mu^a$. 
Thus the indices $I,J$ may include the gauge indices $a,b$. 

General classical solutions of the action (\ref{gauged}) would contain nontrivial flux of the gauge fields and 
two-form field $B_{\mu\nu}$. 
A condensation of tachyons which appear in a constant flux background was discussed recently \cite{flux}. 
The condensation can be understood 
in the context of the gauge 
theory. 
In the following, we will concentrate on solutions which do not contain flux in the non-compact spacetime, i.e. 
we simply set $A^a_\mu,B_{\mu\nu}=0$. 

The condition above requires some of the scalar fields to be constant. 
The scalars which do not couple to the gauge fields are not necessarily constant. 
Unstable solutions with position-dependent 
scalars may have only tachyons localized around a spatial region, and thus we 
assume that all the scalars $\phi^I$ are constant: $\phi^I=\phi_0^I$. 
Then the equations of motion of the scalars require that $\phi_0^I$ correspond to an extremum of 
the scalar potential $V(\phi)$. 

Now the remaining equations are 
\begin{eqnarray}
R_{\mu\nu}+2\nabla_\mu\partial_\nu\Phi &=& 0, \\
4\partial_\mu\Phi\partial^\mu\Phi-2\nabla_\mu\partial^\mu\Phi+V(\phi_0) &=& 0, 
       \label{EOM} \\
\frac{\partial V}{\partial\phi^I}(\phi_0) &=& 0.
\end{eqnarray}
A typical solution to the above equations with $V(\phi_0)\ne0$ is the linear dilaton background 
\begin{eqnarray}
& g_{\mu\nu} = \eta_{\mu\nu}, \hspace{5mm} \Phi = \lambda_\mu x^\mu, & \\
& \mbox{s.t.}\ \ 4\lambda_\mu\lambda^\mu+V(\phi_0) = 0. & \nonumber
\end{eqnarray}
Note that a constant $\Phi$ is allowed to be a solution only when $V(\phi_0)=0$. 

In tachyon condensations, the initial and the final values of the potential should be different,
\begin{equation}
V(\phi_i) > V(\phi_f).
\end{equation}
Therefore, at least one of the background corresponding to UV or IR limit would be a strong coupling background, 
and the worldsheet RG analysis would not be reliable. 

In particular, one can easily make a more strong statement when $V(\phi_0)>0$ and $\Phi$ is static. 
Under those conditions, the dilaton $\Phi$ grows without any upper bound. 
This can be shown as follows: 
Suppose that $\Phi$ is bounded from above. 
Then there is a local maximum of $\Phi$. 
Denote the corresponding point as $P$. 
At $P$, the following equations are satisfied
\begin{equation}
\partial_k\Phi|_P=0, \hspace{5mm} \partial_k\partial_k\Phi|_P<0\ \ (\mbox{no summation for }k),
\end{equation}
where $k$ runs over spatial directions. 
These equations contradict the equation of motion (\ref{EOM}), and thus $\Phi$ must grow without any upper bound. 
This would suggest that a tachyon condensation whose endpoint is the flat supersymmetric vacuum could not be 
approximated by tree-level worldsheet theories.

\vspace{1cm}

\section{The c-theorem}  \label{c}

\vspace{5mm}

In the previous section, we pointed out that the whole processes of bulk tachyon condensations could not be well 
approximated by worldsheet theories due to string loop corrections. 
In this section, we show that bulk tachyon condensations would not conflict with the c-theorem, at least 
within a range 
of the RG flows where string loop corrections are negligible. 

In a worldsheet approach to the off-shell string dynamics, one would assume that a relevant space for string 
theory is the space of all two-dimensional theories, denoted as ${\cal S}$, which can be defined only formally 
\cite{X^2}. 
And the dynamics is assumed to be related to the RG flows in the theory space. 
This prescription seems to apply successfully to open string theory including only boundary perturbations. 
As we have mentioned, this success would be due to the fact that closed string backgrounds can be chosen suitably 
and string loop corrections can be ignored by simply taking the string coupling constant to be small. 
Thus one can concentrate 
only on the degrees of freedom living on some unstable branes. 
On the contrary, in the case of closed string theory the dynamics of the dilaton field must be taken into account. 
Therefore, in that case 
we have to restrict ourselves to a subspace ${\cal S}_{weak}\subset{\cal S}$ in which the 
string coupling is small everywhere in the target spacetime. 

We have discussed non-linear sigma models and their low energy effective theories. 
Thus our interest is in another subspace ${\cal S}'\subset{\cal S}$, where
\begin{eqnarray}
{\cal S}' &=& \{\mbox{space of all non-linear sigma models}\} \nonumber \\
          &=& \{\mbox{space of all configurations of }G_{\mu\nu},\ B_{\mu\nu},\ \Phi,\cdots\}.
\end{eqnarray}
In a subspace ${\cal S}'_{weak}={\cal S}'\cap{\cal S}_{weak}$ in which the string coupling is small, the tree-level 
approximation of string theory is valid. 
For simplicity, we consider a more small subspace ${\cal S}'_{const}\subset{\cal S}'_{weak}$ in which the 
string coupling is a small constant. 

The equation of motion of the dilaton is 
\begin{eqnarray}
0 &=& -R+4\partial_\mu\Phi\partial^\mu\Phi-4\nabla_\mu\partial^\mu\Phi+\frac1{12}H_{\mu\nu\rho}H^{\mu\nu\rho}
        \nonumber \\
  & &+f_{ab}(\phi)F_{\mu\nu}^aF^{b\mu\nu}+g_{IJ}(\phi)D_\mu\phi^ID^\mu\phi^J+V(\phi).
         \label{betafn}
\end{eqnarray}
Since we already set $\Phi=$const., this equation would not be satisfied in general, provided that the other 
equations of motion are satisfied. 
Therefore we discuss an off-shell configuration. 
The corresponding worldsheet theory is no longer conformal. 
However, one can calculate the central charge of such a theory in a special case. 
That is, one can obtain the Virasoro algebra provided that beta-functionals 
of the sigma model coupling functions vanish except for the beta-functional $\beta_\Phi$ of the dilaton. 
In this case, the Virasoro algebra has a central charge $c=\beta_\Phi$. 
Thus we can discuss central charges unless the backgrounds considered do not satisfy all the equations of motion. 

The beta-functional $\beta_\Phi$ is proportional to the right-hand side of the eq.(\ref{betafn}) with positive 
proportionality constant, when 
\begin{equation}
\beta_{G_{\mu\nu}}=\beta_{B_{\mu\nu}}=\cdots=0.
   \label{vanishingbeta}
\end{equation}
For example, a configuration 
\begin{eqnarray}
&& g_{\mu\nu}=\eta_{\mu\nu}, \hspace{5mm} \mbox{no flux}, \nonumber \\ 
&& \phi^I=\phi^I_0, \hspace{5mm} \frac{\partial V}{\partial\phi^I}(\phi_0)=0, 
    \label{config}
\end{eqnarray}
would satisfy the conditions (\ref{vanishingbeta}). 
Note that the scalar potential does not appear in $\beta_{G_{\mu\nu}}$. 
Then the central charge corresponding to this configuration is 
\begin{equation}
c=\beta_\Phi\propto V(\phi_0). 
\end{equation}
Thus the central charge would decrease when the vevs of the scalars go down the potential. 
Note that we have not defined the central charge on whole ${\cal S}'_{const}$, and thus we cannot prove the 
existence of a 
monotonically-decreasing function on that space. 
However, we can say that 
\begin{equation}
c_i>c_f \hspace{5mm} \mbox{if} \hspace{5mm} V(\phi_i)>V(\phi_f). 
\end{equation}
This would suggest the consistency between the physical intuition for bulk tachyon condensations and the 
c-theorem. 

Note that the configuration (\ref{config}) is not a correct background for critical string theory unless 
$V(\phi^I_0)=0$. 
Due to the nonvanishing $\beta_\Phi$, the configuration would be driven to a strong coupling background by the RG 
flow, but analyses outside ${\cal S}'_{weak}$ would be no longer reliable.

\vspace{1cm}

\section{Discussion}   \label{discussion}

\vspace{5mm}

We have discussed bulk tachyon condensations in terms of worldsheet theories whose low energy effective theories 
are of the form (\ref{gauged}). 
Similar discussions could be applicable to more general situations. 
For example, let us consider an effective theory which is written in the Einstein frame 
\begin{equation}
S = \frac1{2\kappa^2}\int d^Dx\sqrt{-g}\left[R+\cdots-V(\phi)\right].
\end{equation}
Suppose that the potential $V(\phi)$ does not depend on the dilaton $\Phi$ so that one can take an arbitrarily 
small value of the string coupling constant to ensure the validity of the tree-level approximation. 
Then in the string frame, the corresponding action is
\begin{equation}
S = \frac1{2\kappa^2}\int d^Dx\sqrt{-\tilde{g}}e^{-2\Phi}\left[\tilde{R}+\cdots-e^{-\frac4{D-2}\Phi}V(\phi)\right]. 
\end{equation}
Now the potential depends on the dilaton $\Phi$, and the expression of the central charge 
\begin{equation}
c=\beta_\Phi \propto V(\phi)+\cdots
\end{equation}
with positive proportionality constant would still hold when $\Phi$ is constant. 
Therefore, the consistency with the c-theorem would be independent of the choice of the effective theory 
(\ref{gauged}). 

We have argued that quantum corrections may become important in bulk tachyon condensations. 
A naive expectation for effects of such quantum corrections is the modification of the scalar potential, or more 
specifically, the addition of a quantumly-induced cosmological constant to the scalar potential. 
Thus, even if at the tree-level there are bulk tachyons and vanishing cosmological constant (e.g. in the case of 
various non-supersymmetric string theories), 
the quantum-corrected background might be nontrivial, and the vacuum energy would 
be larger than zero. 
Then, there might be a 
possibility for such a background to decay into a simple symmetric stable vacuum like the flat spacetime 
rather than into a complicated 
non-critical string theory. 

According to our strategy, 
the most appropriate setup to discuss the tachyon condensation would be the case in which the low energy 
effective theory is a gauged supergravities which can have a nontrivial scalar potential without spoiling 
spacetime supersymmetry. 
The scalar potential is determined by the supersymmetry. 
If one finds a local maximum or saddle of the potential, one could find a string theory with bulk tachyons. 
The fate of the unstable theory could be deduced by finding a more stable extremum of the potential. 
If this extremum is the (local) minimum, it is the endpoint of the tachyon condensation. 
If not, the theory continues to decay. 

The possible endpoints of the condensations, however, would not necessarily be ordinary compactifications of 
superstring theories. 
As discussed before, one would have to face a strong coupling dynamics. 
Moreover, there are other issues which would be interesting to investigate further. 

One of them is on the physical interpretation of the extrema of the potential. 
The scalars in the effective theory would include some components of the metric of the internal space. 
For the internal space to be regular, the internal metric must have its non-vanishing determinant. 
However, that condition would not always be satisfied at every extrema of the potential, and thus the corresponding 
background would be a singular compactification. 
The appearance of such a decay would be naively expected according to the c-theorem, 
although the endpoint of the condensation in the case would be difficult to be interpreted. 
Roughly speaking, relevant deformations decrease the degrees of freedom of the theory considered. 
And in particular, unstable sigma models with compact target spaces would be expected to decay into trivial theory 
in the 
infrared. 
Our arguments in this paper do not exclude such possibilities; we could give more detailed discussions to 
the situations. 

Another issue is on the nature of scalar potentials in supergravities. 
In supersymmetric field theories without gravity, the scalar potential is always positive definite. 
Thus supersymmetric vacua are the most stable states, and a non-supersymmetric excited state would probably decay 
into one of the supersymmetric vacua unless there is an obstruction. 
On the other hand, supergravities can have scalar potentials which are not bounded from below. 
A simple example is \cite{unbounded}
\begin{equation}
V(\phi) = \Lambda-a(\cosh\phi-1).
\end{equation}
This potential does not have even a local minimum. 
The scalar $\phi$ is tachyonic, and its condensation would not have any endpoint, in the viewpoint of  the 
tree-level analysis. 
In gauged supergravities, one usually discusses, not vacuum solutions, but domain wall solutions which can preserve 
several supercharges. 
Therefore a possible endpoint of the decay in a condensation in the case 
with an unbounded potential might be such a position-dependent 
stable configuration. 
It would be interesting if this kind of processes can be understood as 'dynamical formations' of 
lower-dimensional braneworlds embedded in a ten-dimensional target spacetime. 

An important lesson of our investigation is that the analysis of 
whole processes of bulk tachyon condensations would be beyond 
the reach of the string perturbation theory. 
Therefore, the use of the closed string field theory might not be enough to handle the problem, 
in contrast with the open string tachyon condensations, 
and it would be necessary to employ some other powerful ingredients.

\end{document}